\documentclass[11pt,twoside]{article}

\usepackage{asp2006}
\usepackage{epsf}
\usepackage{graphicx}
\usepackage{lscape}

\newcommand{\HII}{H{\sc ii}}

\begin{document}
\title{Probing the Early Evolution of Young High-Mass Stars}   

\author{E.~Puga\altaffilmark{1}, A.~Bik\altaffilmark{2}, L.~B.~M.~F.~Waters\altaffilmark{3}, Th.~Henning\altaffilmark{4}, L.~Kaper\altaffilmark{3}, M.~van~den~Ancker\altaffilmark{2},  A.~Lenorzer\altaffilmark{5},  E.~Churchwell\altaffilmark{6}, S.~Kurtz\altaffilmark{7}, J.~A.~Rod\'on \altaffilmark{4}, T.~Vasyunina\altaffilmark{4},  M.~B.~N.~Kouwenhoven\altaffilmark{8}, H.~Beuther\altaffilmark{4}, H.~Linz\altaffilmark{4},  M.~Horrobin\altaffilmark{3}, A.~Stolte\altaffilmark{9}, A.~de Koter\altaffilmark{3}, W.~F.~Thi\altaffilmark{10}, N.~L.~Mart{\'{i}}n-Hern\'{a}ndez\altaffilmark{5}, B.~Acke\altaffilmark{1}, F.~Comeron\altaffilmark{2}, G.~van~der~Plas\altaffilmark{2}, Ch.~Waelkens\altaffilmark{1}, C.~Dominik\altaffilmark{3}, M.~Feldt\altaffilmark{5}}   

\altaffiltext{1}{Instituut voor Sterrenkunde, Katholieke Universiteit Leuven, Celestijnenlaan 200D, B-3001 Leuven, Belgium}
\altaffiltext{2}{European Southern Observatory, Karl-Schwarzschild Strasse 2, Garching-bei-M\"unchen, D-85748, Germany}
\altaffiltext{3}{Astronomical Institute "Anton Pannekoek", University of Amsterdam, Kruislaan 403, 1098 SJ Amsterdam, The Netherlands}
\altaffiltext{4}{Max-Planck-Institut f\"ur Astronomie, K\"onigstuhl 17, D-69117 Heidelberg, Germany}   \altaffiltext{5}{Instituto Astrof{\'{i}}sico de Canarias, Via L\'actea s/n, E-38200 La Laguna, Spain}
\altaffiltext{6}{University of Wisconsin,  475 North Charter Street, WI 53706 Madison, USA}
\altaffiltext{7}{Instituto de Astronom{\'{i}}a UNAM-Morelia, Aptdo. 3-72, 58090 Morelia, Michoac\'an Mexico}
\altaffiltext{8}{University of Sheffield, Hounsfield Road,  Sheffield S3 7RH,  United Kingdom}
\altaffiltext{9}{Department of Astronomy, UCLA, 475 Portola Plaze, Los Angeles, CA 90095, USA}
\altaffiltext{10}{Royal Observatory of Edinburgh, Blackford Hill, Edinburgh EH9 3HJ, UK}


\begin{abstract} 
Near-infrared imaging surveys of high-mass star-forming regions reveal an amazingly complex interplay between star formation and the environment (Churchwell et al. 2006; Alvarez et al. 2004). By means of near-IR spectroscopy the embedded massive young stars can be characterized and placed in the context of their birth site. However, so far spectroscopic surveys have been hopelessly incomplete, hampering any systematic study of these very young massive stars. New integral field instrumentation available at ESO has opened the possibility to take a huge step forward by obtaining a full spectral inventory of the youngest massive stellar populations in star-forming regions currently accessible. Simultaneously, the analysis of the extended emission allows the characterization of the environmental conditions. The Formation and Early Evolution of Massive Stars (FEMS) collaboration aims at setting up a large observing campaign to obtain a full census of the stellar content, ionized material, outflows and PDR's over a sample of regions that cover a large parameter space. Complementary radio, mm and infrared observations will be used for the characterization of the deeply embedded population. For the first eight regions we have obtained 40 hours of SINFONI observations. In this  contribution, we present the first results on three regions that illustrate the potential of this strategy. 
\end{abstract}


\section{Introduction}
High-mass stars form quickly, while deeply embedded in molecular clouds. These sites are only accessible at far-infrared and (sub)mm wavelengths. However, in  these wavelength regimes, only indirect information about the forming stars can be derived.  When the stars  form, they start to destroy the surrounding molecular gas cloud and  become visible at shorter wavelengths. The near-IR atmospheric  window offers 
the unique possibility to access the photosphere of the stellar 
component -- peering through the typical obscuring environment -- while the diffuse cold dust emission does not yet dominate.\\
This enables us to study the  fundamental stellar parameters of the young massive stars and  the interaction between the young stars and their environments. Additionally, these regions are  young enough  to search for relics of the star formation process by means of their infrared excess (Bik et al. this volume).\\

Observations show that high-mass stars preferentially form in clusters (de Wit et al. 2005). Hence, if we wish to understand high-mass star formation in general, we must understand  the \emph{clustered} mode of star formation.
Supporting evidence of this hypothesis comes from near-IR surveys that have unveiled  the presence of young high-mass  clusters close to UC\HII{} regions (Hanson et al. 2002; Alvarez et al. 2004; Bik PhD Thesis 2004; Kaper et al. 2008 in prep.). These surveys have provided  a valuable view on the complex morphology of these regions via imaging. From such studies we learn that the detected embedded clusters differ in morphology between very compact and related to UC\HII{} regions, to very extended clusters with only diffuse \HII{} emission.  \\
Follow-up long-slit spectroscopy of a few selected objects (typically 5 of up to 100 or more stars in each field) has allowed the classification of a few selected sources in these clusters.  The diversity of  properties of our selected objects spans between: (1) "Naked" OB stars with photospheric spectra (Bik et al. 2005); (2) Massive Young Stellar Objects (Blum et al. 2004; Bik et al. 2006) which show a large variety
  of spectral properties in the circumstellar material; (3)  Near-infrared point-sources which are the counterparts to the radio UC\HII{} regions (Mart{\'{i}}n-Hern\'andez et al. 2003).\\
This limited number of $K-$band spectra already demonstrates the
tremendous power and diagnostic value of near-IR spectroscopy in
characterizing the young stellar population. However, the complexity of the fields and poor
  spectroscopic sampling of the population prevents any systematic
  investigation of the stellar content in the context of their
  environment. \\

Two important questions still remain to be answered:  1) What is the high-mass stellar census of the young stellar  clusters and how does this depend on the cluster properties (such as
  luminosity and evolutionary phase)? 2) What is the interaction between  the massive stars and their immediate environment? \\ 
The recent development of near-IR integral-field spectroscopy  offers a great opportunity since it provides  simultaneous and complete information on the stellar content and on the diffuse nebular emission. 

\section{The FEMS Project}

The FEMS project comprises
a plethora of aspects relevant to  high-mass star formation and evolution processes.
With our multi-disciplinary approach we will be able to classify the stellar content and derive the distribution of OB stars, massive young stellar objects (YSOs) and
the lower-mass pre-main sequence stars. Detailed stellar parameters of the OB stars (e.g., $T_{\rm{eff}}$, log $g$, $v$ sin $i$) will be derived by fitting the
spectra using state-of-the-art stellar atmospheric models  like FASTWIND (Puls et al. 2005) and  CMFGEN (Hillier \& Miller, 1998). The OB stellar population also provides  an
independent distance estimate by using the spectroscopic parallax of  late-O/early-B stars. For the very rich clusters we can derive a reliable initial mass function (IMF) by having a
spectroscopic classification of the high-mass end.\\
The extended emission present in these regions in the vicinity of those 
massive stars can be utilized to infer extinction properties, and to provide 
details about possible outflows as well as the surrounding emission nebulae 
(photon-dominated regions or PDRs  and H{\sc ii} regions).
This enables an in-depth study of the interplay
between the stellar content and its surrounding medium as well as the influence of photo-evaporation and outflows.\\
The FEMS project also follows a multi-wavelength
approach, since additional longer wavelength data in the mid-IR,
(sub)mm and radio range will unveil the extremely embedded and possibly even
younger regions of star formation as well as large scale outflows. In the end, this
large homogeneous data set will give an unbiased census of all the young high-mass
stars and their environment which is needed to construct a full picture of
embedded clusters.

\section{First Results}
The first SINFONI observations for the FEMS project were performed between February and April 2007 on 8 complete and 3 partly covered high-mass star-forming regions to
demonstrate the feasibility of the projects as envisaged above.   The observations were performed with SINFONI in the H+K setting (R=1500) using a 1/2-overlap raster technique to mosaic a typical area of 50$\arcsec\times$50$\arcsec$.\\
In the following sections, we describe a few preliminary studies that have been undertaken with these SINFONI data on  three representative objects, included in our sample.

\subsection{Evolved Clusters: IRAS 08563-4711}

The region IRAS 08563-4711 coincides with the location of an \HII{} region in the Vela Molecular Ridge, at a distance of 2 kpc. The near-IR view of this object shows  a stellar cluster (Bik PhD Thesis, 2004) obscured by an average extinction $A_{\rm{V}}$ $\sim$ 15 mag.\\
From our SINFONI data, 73 stellar spectra were extracted with a SNR between 5 and 120.
The stellar content of our cluster has been classified on the basis of its emission and/or absorption lines (Hanson et al. 1996). With this direct approach, we are able to immediately identify up to a mid-B spectral type (O9V and four B-type stars), as well as a late-type star (see Fig.~1). \\

\begin{figure}[!ht]
\begin{center}
\includegraphics[width=13cm, scale=.4]{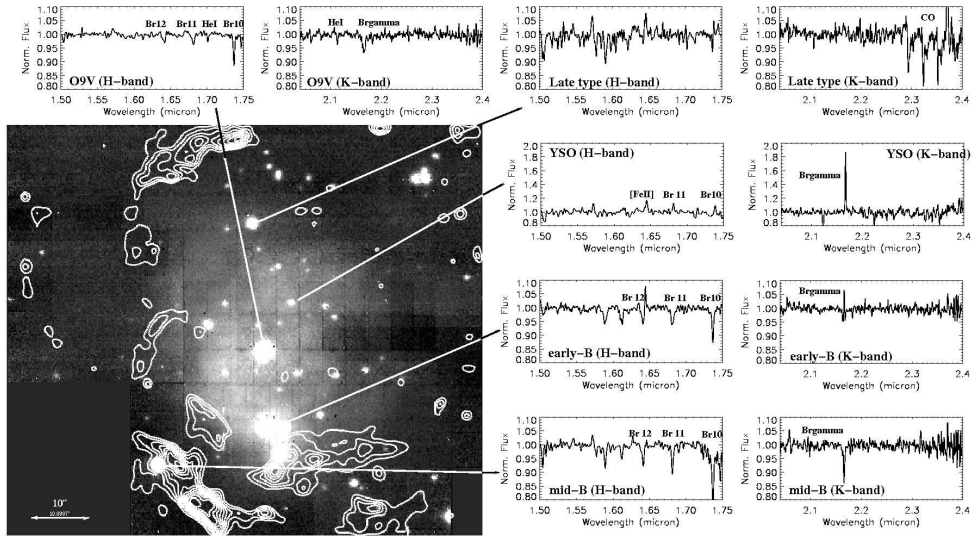}
\caption{Image of the Br$\gamma$+continuum emission toward IRAS 08563-4711. Overlayed contours represent the H$_2$ 1--0 S(1) line that encompasses the Br$\gamma$ emission, likely tracing the PDR. We show the spectra of several representative examples of the various objects found among its stellar population.  }
\end{center}
\end{figure}

Combining  the spectral classes found with the color-magnitude diagram (CMD) (Kaper et al. 2008) we find that the brightest star in the cluster is the ionizing O9 star (see Fig.~2 {\it Left}). Furthermore, the spectrophotometric distance agrees with the distance derived from $V_{\rm{LSR}}$ measurements of CS (Bronfman 1996).\\
The B-stars are spread out over an enormous range of extinctions, ranging from $A_{\rm{V}}$ $=$ 3--23 mag, reflecting the fact that some of them are
located in the outskirts, where the extinction is higher due to remnant blobs of dense material not yet dispersed by the inner 
cluster stars. This CMD  shows that we can access the high-mass stellar populations up to  $A_{\rm{V}}$'s $\sim$ 30 mag for the O stars while the low-mass pre-main sequence population in the H{\sc ii} region is also detected.\\
Determination of the spectroscopic  mass via stellar atmospheric models of OB stars will  contribute to the construction of a more reliable IMF for the high-mass stars.\\
\begin{figure}[!ht]\label{fig_2}
\begin{center}
\begin{minipage}[c]{.5\textwidth}\includegraphics[width=5cm, scale=0.6]{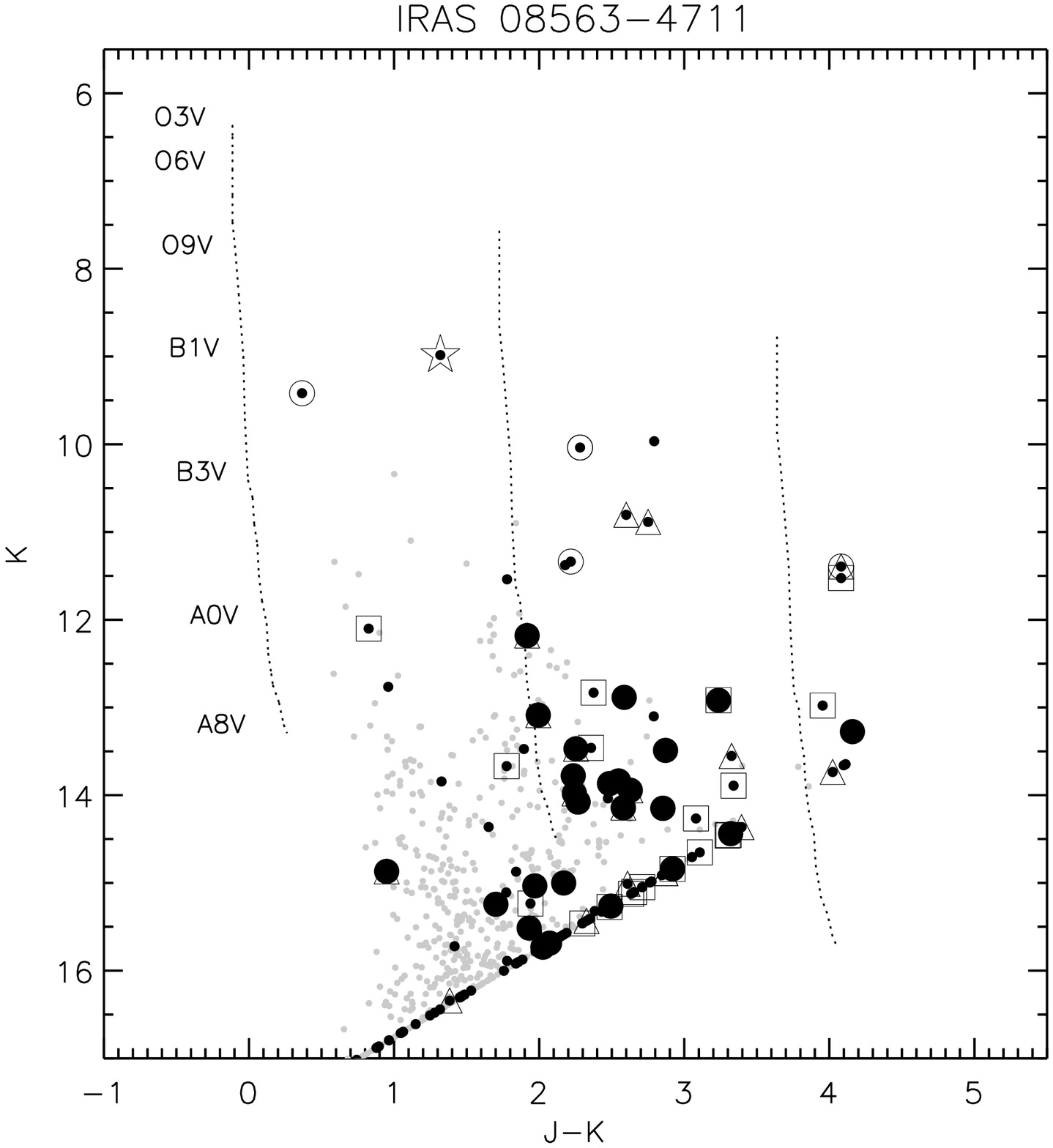}\end{minipage}\hspace{-1.5cm}\begin{minipage}[c]{.5\textwidth}\includegraphics[width=5.5cm, scale=0.3, angle=270, bb= 92 152 510 694]{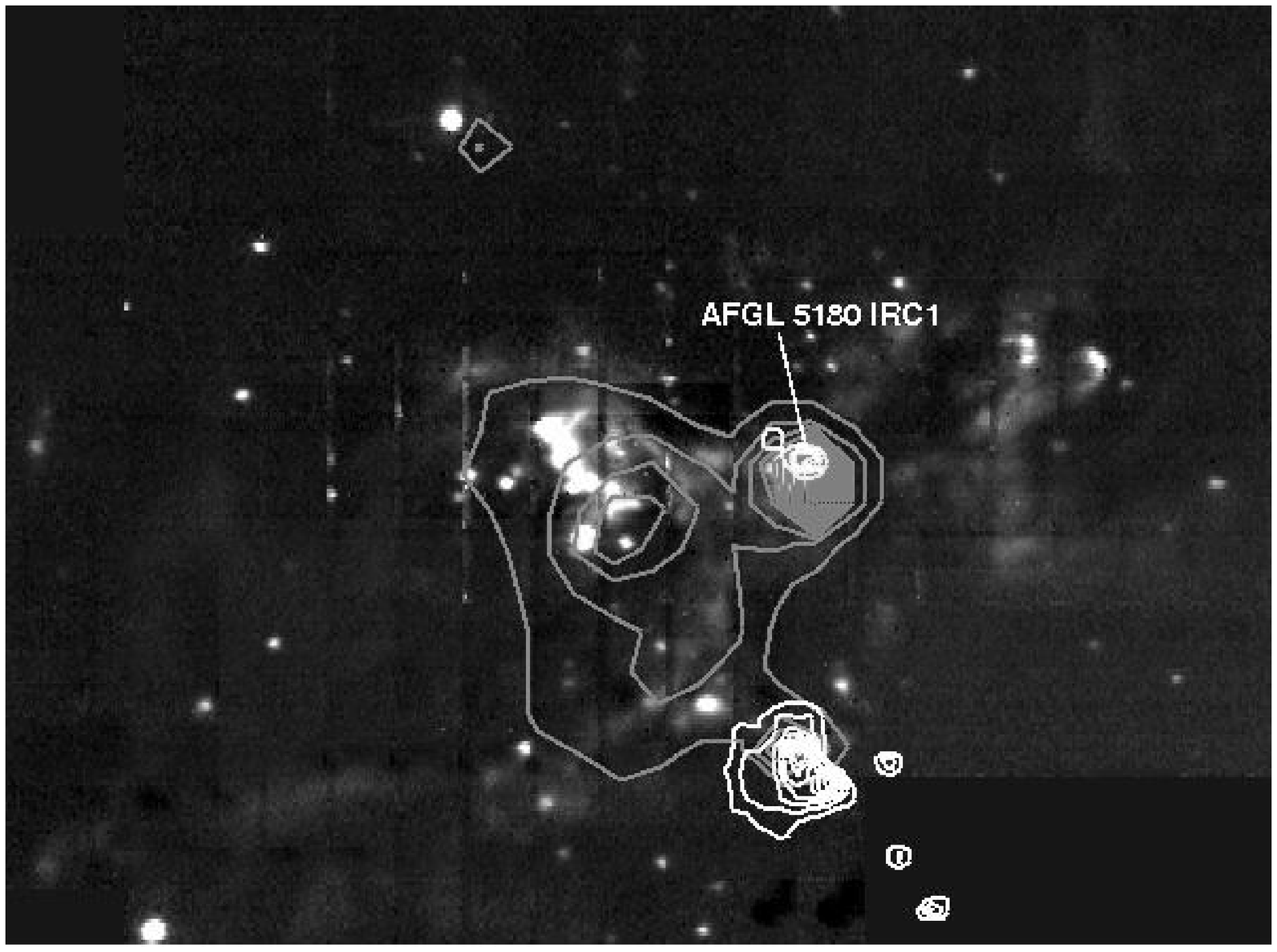}\end{minipage}

\caption{ {\it Left:} CMD of the stellar population around IRAS 08563-4711 obtained with NTT photometry. Our SINFONI spectroscopic classification has been added to the diagram ({\it star}: O star, {\it open circle}: B-star, {\it filled circle}: YSO, {\it triangle}: nebular and {\it open square}: featureless spectra). {\it Right:} Multi-wavelength view of IRAS 06058+2135. The gray scaled image is the emission in the H$_2$ line as observed by SINFONI, the gray contours represent the 8 $\mu$m image taken with SPITZER and the white contours are 1.2 mm SMA observations. The mm data reveal a different, and  probably much younger, regions of star formation to the south-west of the cluster detected in the near-IR, suggesting multiple epochs of star formation.}
\end{center}
\end{figure}
The extended nebular emission detected by our SINFONI data toward IRAS 08563-4711 traces a classic \HII{} region in the Brackett $\gamma$, 10--4 and 9--4 lines. We also detect HeI emission at 2.058 $\mu$m in the close proximity of the central massive star.  Extended  H$_2$ 1--0 S(1) emission encloses the \HII{} region in a shell structure, tracing the  PDR.

\subsection{Multiple Epochs of  Star Formation: IRAS 06058+2138 (AFGL 5180, S247)}

The combination of  SINFONI data with SPITZER and mm data is a very
powerful way to detect different stages of star formation.
The case of  IRAS 06058+2138 illustrates this  multi-epoch scenario  (see Fig.~2 {\it Right}). The near-IR data show the
less embedded cluster to the west. The SPITZER 8 $\mu$m image (grey contours) shows
a very bright source next to the cluster, while the SMA data (white
contours) reveal a likely much younger cluster of objects  clearly offset from the near-IR cluster (Rodon et al. in prep.).

H$_2$ and [FeII] emission are detected in the surroundings of a bright red object in the field, AFGL 5180 IRS1 (Longmore et al. 2006). These resolved structures exhibit a bow-like shape that clearly trail back to the infrared source, indicating a very young age.

\subsection{Importance of the Outflow Activity:  IRAS 06084-0611 (GGD 12-15)}

IRAS 06084-0611 is associated with the cometary UC\HII{} region G213.880-11.837 (Gomez, Rodriguez \& Garay, 2002), located at a  distance of 1 kpc in the Gem OB1 complex. A cluster of red sources appears  in the proximity of the UC\HII{} region, albeit no near-IR counterpart is detected at the location of the radio continuum source. Two objects with strong near-IR excess were considered to be massive YSO candidates through long-slit spectroscopic studies in the K-band (Bik et al. 2006). Persi \& Tapia (2003) concluded from mid-IR data that the luminosity budget of the southern source (nr118) corresponds to a spectral type later than B3.\\
Our preliminary analysis of the SINFONI data on point-like sources shows that this last YSO candidate has a  very rich hydrogen recombination line spectrum (Brackett and Pfund series) indicating hot ionized hydrogen. This suggests the presence of  hot dense material close to the central star.

The northern YSO candidate (nr114) does not show radio continuum emission in  VLA observations (Gomez, Rodriguez \& Garay, 2002). Meanwhile, the H$_2$ v=1-0\,S(1) mosaic map  of this region shows very prominent unresolved line emission at three locations (See Fig.~3). Two of these emission blobs are approximately equidistant from the  massive YSO candidate nr114, located at the north of the image and oriented in  east-west direction.\\
 The emission of molecular hydrogen can be caused by  heating from shocks. Alternatively, it could be produced by fluorescence excitation by non-ionising UV photons (i.e., in PDRs). Especially in dense clouds, UV irradiation can mimic the H$_2$ spectra produced by   emission due to shocks. In order to help distinguish between these two mechanisms, the line ratios of several ro-vibrational H$_2$ lines can be compared.

\begin{figure}[!t]\label{fig_3}
\begin{center}
\includegraphics[scale=0.7, angle=0]{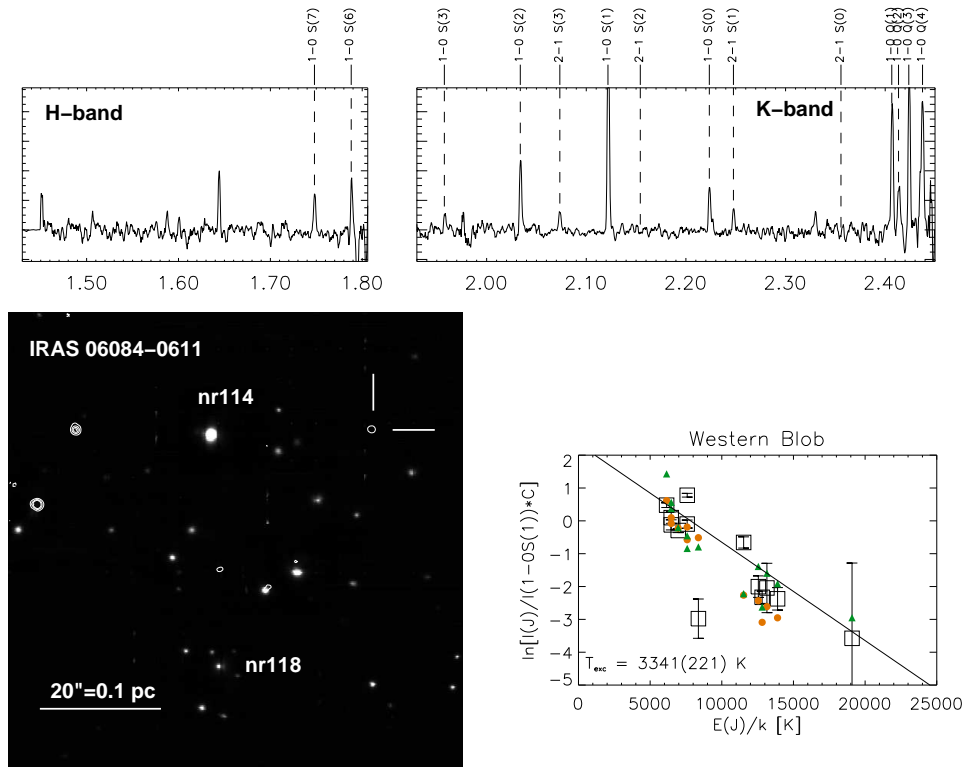}
\caption{({\it Bottom-left}) Br$\gamma$+continuum map around IRAS 06084-0611. Contours represent the H$_2$ v=1-0 S(1) line emission. Two of the prominent features appear equidistant with respect to the candidate YSO  nr114. ({\it Top}) Extracted spectrum of the western blob. A $A_{\rm{V}}$ $=$ 7 mag is derived from the ratio between v=1-0 {S(1), Q(3)} and {S(0), Q(2)}. ({\it Bottom-right}) Excitation diagram of the ro-vibrational lines and comparison with (circles) LTE model  and (triangles) fluorescence emission model with n=10$^6$ cm$^{-3}$ from Draine \& Bertoldi (1996).}
\end{center}
\end{figure}  

Detection of the Q series allows an estimate of the extinction toward these blobs from pairs of lines arising from the same upper level v=1-0 {S(1), Q(3)}, {S(0), Q(2)} (Beckwith et al. 1979). 
These ratios render an approximate visual extinction of 7 mag. In our excitation diagrams, we compare the (unreddened) measured line ratios of several lines to the predictions of both an LTE model at 2000K and a high-density photodissociation region  model (Draine \& Bertoldi, 1996). The measured line ratios are well fit by a purely thermal single-temperature model of  around 3000K. The low extinction measured suggests that a high-density PDR and a strong UV field can be ruled out in favor of a shock  scenario.\\
The confirmation of the shock nature of the line emission associated to nr114 therefore reinforces the hypothesis that this source is extremely young.

\section{Summary and Conclusions}

We have demonstrated the power of near-IR integral field observations  of a few high-mass star-forming regions. We have explored the utility of these observations to characterize the stellar content of a not-so-young cluster, the line emission associated with very young stars and the multiple-epoch formation scenario in a third region.
Our results show that complementary information on the stellar content and the extended nebular emission in the near-IR can substantially improve our understanding of high-mass star formation.\\
We  conclude that it is now possible
to take a huge step forward by obtaining a full spectral inventory
 of the youngest massive stellar populations in star-forming regions. Furthermore, we can also study the interplay between the cluster and  the surrounding environment through the analysis of the nebulae's physical properties.  

\acknowledgements 
 E. Puga would like to thank the organizers of the MSF07 Conference for the support provided to attend this meeting. 


\end{document}